\documentclass[]{aa}

\usepackage[dvips]{graphicx}
\usepackage{amsmath}
\usepackage{amssymb}

\begin{document}
\title{Optical and Near-IR Spectra of O-rich Mira Variables :
a Comparison between Models and Observations}
\author{A. Tej\inst{1}, A. Lan\c{c}on\inst{1}, M. Scholz\inst{2}
 	\& P. R. Wood\inst{3}}

\offprints{A. Lan\c{c}on}

\institute{
 Observatoire Astronomique de Strasbourg,
 Universit\'e L.\,Pasteur \& CNRS (UMR 7550), Strasbourg,
France;\\
 \email{surname@astro.u-strasbg.fr}
\and
Institut f. Theoretische Astrophysik der Universit{\"a}t Heidelberg,
Tiergartenstr.15, 69121 Heidelberg, Germany, and School of Physics,
University of Sydney, NSW 2006, Australia;\\
 \email{scholz@ita.uni-heidelberg.de}
\and
Research School for Astronomy and Astrophysics,
Australian National University, Canberra ACT 2600, Australia;
 \email{wood@mso.anu.edu.au}
}

\date{Received February 2003 / Accepted September 2003}

\authorrunning{A. Tej et al.}
\titlerunning{Comparison of Mira models with observations}
\abstract{
Pulsation models are crucial for the
interpretation of spectrophotometric and interferometric
observations of Mira variables. Comparing predicted and observed
spectra is one way of establishing the validity
of such models. In this paper, we focus
on the models published between 1996 and 1998 by Bessell,
Hofmann, Scholz and Wood. A few new model spectra are
added, to improve available phase coverage. We compare the synthetic
spectra with observed low resolution spectra of optically selected
oxygen-rich Miras, over a range of optical and near-IR 
wavelengths that encompasses most of the stellar energy output.
We investigate the overall energy distributions, and specific
spectral features in the near-IR wavelength range.
The agreement between the observed and the model-predicted
properties is found to be reasonably good. However, there are
discrepancies seen especially in various molecular bands. \\
We find that different combinations of stellar parameters and 
pulsation phases often result in very similar model spectra.
%(differences between two such spectra are smaller than typical
%differences between an observed spectrum and its best fit model).
Therefore the problem of deriving parameters of a Mira variable from 
its spectrum has no unique solution. More advanced models than presently 
available, providing even better fits to the data and covering a wider range 
of parameters, would be needed to achieve better discrimination. 
}
\maketitle
\section{Introduction}
\label{intro.sec}
Mira variables play a major role in understanding
stellar evolution and galactic spectral evolution. 
Stellar lifetimes, the contribution of AGB stars to the integrated light 
emission of stellar populations, or the chemical yields of intermediate 
mass stars depend critically on properties such as stellar mass, pulsation
and mass loss. Hence, estimates of the fundamental stellar parameters
and atmospheric properties of these stars, from
spectroscopic, photometric and interferometric observations,
remain crucial. The interpretation of any of these data  
relies on stellar models.

In a previous paper (Tej et al. 2003 = TLS03), based on models of Bessell
et al. (1996 = BSW96) and Hofmann et al. (1998 = HSW98), we have shown
 that simultaneous spectrophotometric observations of
absorption bands of H$_{2}$O 
and interferometric measurements in
the H$_{2}$O-contaminated near-continuum filters in the near-IR may 
be useful in deriving physically sensible diameters from the 
uniform-disk fitted interferometric values. 
The mentioned set of models is widely used in interpreting
interferometric measurements (e.g. Perrin et al. 1999; Mennesson et al. 2002).
However, the atmospheres of long period variables
(LPVs) are particularly complex due to their extended
and dynamical configuration (e.g. Scholz 2003). 
Despite continuing progress 
(e.g. Bowen 1988; H{\"o}fner \& Dorfi 1997; H{\"o}fner et al. 2003), 
completely self-consistent models for the
time-dependent stellar stratifications and spectra do not exist.
Simplifying assumptions in the radiative transfer codes make the
predictions intrinsically uncertain. Hence, confronting the
model-predicted properties with observed data is fundamental for
checking the validity of the models. Comparison between predicted
and observed spectra provides a direct and sensitive way of doing
this.

In this paper, we focus on the optical and near-IR emission of
oxygen-rich LPVs, as observed by Lan\c{c}on \& Wood (2000 = LW2000)
and predicted by the models of BSW96 and HSW98. 
The selected data consists of low to medium 
resolution spectra from 0.5 to 2.5\,$\mu$m, for stars
with a range of pulsation properties and estimated masses and 
metallicities. This is the first time a comparison over such an
extended range of wavelengths is carried out with dynamical models. 
For the carbon stars of LW2000 with small variation amplitudes, 
a comparative study has been done by Loidl et al. (2001). 
They show that the observed spectra can be reproduced satisfactorily
with hydrostatic model
atmospheres based on the MARCS code (Gustafsson et al. 1975;
J{\o}rgensen 1997). For O-rich LPVs, 
most of the recent comparison work has been devoted to 
the longer infrared wavelengths observed with the 
ISO\footnote{Infrared Space Observatory} satellite.
Aringer et al. (2002) and Matsuura et al. (2002) have studied
the ISO/SWS spectra covering the wavelength range from 2.36 --
5.0\,$\mu$m and 2.5 -- 4.0\,$\mu$m respectively.
These studies 
focus on the combined effects of molecular bands, but
their spectral coverage is not sufficient to address 
the molecular features and the energy distribution of
the stellar emission simultaneously. For Miras, they illustrate
the limits of hydrostatic models nevertheless.
Alvarez et al. (2000) showed that
the version of the static MARCS models described by Bessell
et al. (1998) and Alvarez \& Plez (1998) is capable of reproducing the
optical spectra of many variable M giants. But
agreement is lost when the optical and near-IR spectra are 
studied jointly, and a systematic study of the complete
spectra was postponed by this group to times when pulsating models would
become available.

The aim of this paper is not to analyse in detail any individual
observed star. The large number of relevant stellar parameters
(mass, effective temperature, luminosity, abundance ratios
in the atmosphere, pulsation mode, period and amplitude, 
pulsation phase {\em and} pulsation cycle, circumstellar extinction, etc.) 
and the limited number of available models makes it unlikely to find a 
perfect physical match. 
The probability of finding a matching time series of models
for a series of repeated observations of an individual star
is essentially zero. Our purpose is therefore mainly to compare 
global properties and trends. 

In Sect.\,\ref{mod_data.sec} we summarize the 
properties of the adopted models and discuss relevant details of
the observed spectra. In Sect.\,\ref{compare.sec}, we show that
the global shape of the models is satisfactory, in that 
a variety of observed spectra can be reasonably reproduced. 
In Sect.\,\ref{features.sec} we focus on selected spectral 
features which includes the broad-band colours, the 1\,$\mu$m 
slope and the molecular bands. In that section, we identify and 
discuss discrepancies. We conclude in Sect.\,\ref{disc_con.sec}.

%------------------------------------------------------------------
\section{Models and Observation}
\label{mod_data.sec}
\subsection{Model Spectra}
\label{models.sec}

The physical assumptions of the nonlinear pulsation models
that we investigate are described in BSW96 and HSW98. We recall the
relevant and important aspects here and refer the reader to these
papers for the detailed description.
The HSW98 models are completely self-excited configurations whereas, for
the BSW96 models, pulsation of the atmospheric outer layers are produced by
applying a conventional piston to the sub-atmospheric layers. For similar
parent star parameters the HSW98 models show substantially more extended
atmospheres and display stronger cycle-to-cycle variations. Non-grey
temperatures are computed in the approximations of local thermodynamic and
radiative equilibrium, i.e. instantaneous relaxation of
shock-heated material behind the shock front is assumed. 
The density stratifications are determined from
shock-front driven outflows and subsequent infall of matter. Details of opacity
contributions and numerical treatment of molecular band absorptions are
outlined in BSW96, HSW98 and Brett (1990). For H$_{2}$O, the empirical
absorption coefficients of Ludwig (1971) were used.

For convenience,
Table \ref{MiraParent.tab} recalls the parameters of the (static)
parent stars of the pulsating models and the bolometric amplitudes
of pulsation. All the models are constructed for
masses of 1 to 2 M$_{\odot}$ and solar abundances. 
Because they were designed to
reproduce the physical properties of the prototype Mira variables
$o$ Ceti and R Leo, they have periods $\sim$ 310 -- 330 days, close
to the periods of these two stars.
Parameters of individual models at different phases are
summarized in the Appendix. Due mainly to irregularities
and cycle-to-cycle variations of the light curves, phase zero points are
affected by a small degree of arbitrariness (see also
Sect.\,\ref{compare.sec}). Besides Rosseland radii $R$
and corresponding effective temperatures $T_{\rm eff}$,
we also give near-continuum radii $R_{1.04}$ 
and corresponding effective temperatures $T_{\rm 1.04}$.
Unity is reached by the mean Rosseland optical depth at $R$, 
and by the monochromatic optical depth at 1.04\,$\mu$m at $R_{1.04}$.
$R_{1.04}$ is not affected by problems related to strong high-layer
molecular absorption that enter the Rosseland
opacity (see, e.g., HSW98, Scholz 2003): significant effects may
be found below about 2800K, in particular for the very extended P series
atmospheres where Rosseland effective temperatures may be up to several hundred
degrees lower than 1.04 temperatures. We will use $R_{1.04}$ and $T_{\rm 1.04}$
values as reference quantities in this paper.
Within a series, the phase variations of effective temperature of several 
hundred degrees are typical\,;
the effective temperatures of all models range from about 2100\,K
(O series near minimum light) to about 3300\,K (Z series near maximum
light) and still higher values (around 3700\,K) found at pre-maximum phases of
fundamental-mode pulsation models (see discussion in the Appendix).
\begin{table*}
\caption{Properties of the Mira `parent' star series.
The columns: pulsation mode - fundamental (f)
or overtone (o); period $P$; parent star mass $M$; luminosity
$L$; Rosseland radius $R_{\rm p}$; effective temperature $T_{\rm eff}
\propto\, (L/R_{\rm p}\,^{2})^{1/4}$ and bolometric amplitude $\Delta
M_{\rm bol}$.}
\label{MiraParent.tab}
\begin{center}
\begin{tabular}{lllllllll}
\hline
Series & Mode & $P$ & $M$
 & $L$ & $R_{\rm p}$ & $T_{\rm eff}$ & 
 $\Delta M_{\rm bol}$ & Reference\\
 & & (days) & (M$_{\odot})$ & (L$_{\odot})$ & (R$_{\odot})$ & & \\
\hline
Z & f & 334 & 1.0 & 6310 & 236 & 3370 & 1.0 & BSW96 \\
D & f & 330 & 1.0 & 3470 & 236 & 2900 & 1.0 & BSW96 \\
E & o & 328 & 1.0 & 6310 & 366 & 2700 & 0.7 & BSW96 \\
P & f & 332 & 1.0 & 3470 & 241 & 2860 & 1.3 & HSW98  \\
M & f & 332 & 1.2 & 3470 & 260 & 2750 & 1.2 & HSW98  \\
O & o & 320 & 2.0 & 5830 & 503 & 2250 & 0.5 & HSW98  \\
\hline
\end{tabular}
\end{center}
\end{table*}

The spectra of Miras vary considerably with phase, 
and the variations of individual spectral features are
not in phase with each other (Alvarez \& Plez 1998). 
Hence, a comparative study requires that the models
be considered not only at minimum and maximum light.
In this paper we have included new phase 0.2 and 0.8 models 
of the P and M series. 
These models were not discussed in HSW98. Their parameters and
a brief description are included in the Appendix. 

Taken as a sample, the models display a wide variety of atmospheric 
structures in terms of density, temperature and
molecular abundance stratifications (e.g. TLS03). 
As a result, they also produce a broad distribution of spectral
properties.
It must however be kept in mind that the models cover a 
limited range of parameters (e.g. in terms of period, amplitude,
metallicity, mass).
It is currently not possible to perform an extensive analysis of
the parameter-dependences of the model spectra: this will
require a larger systematic model set.

\begin{figure}
\includegraphics[clip=,width=0.5\textwidth]{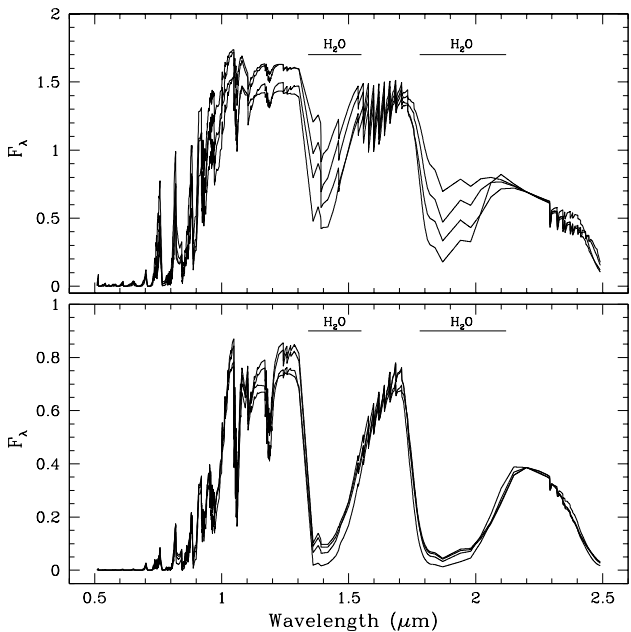}
\caption{This figure shows shape of the H$_2$O bands for the
near-maximum (upper panel) and the near-minimum (lower panel)
P models. The flux density per unit wavelength
is in arbitrary units and normalized at 2.2\,$\mu$m.}
\label{shape.fig}
\end{figure}

\begin{figure}
\includegraphics[clip=,width=0.5\textwidth]{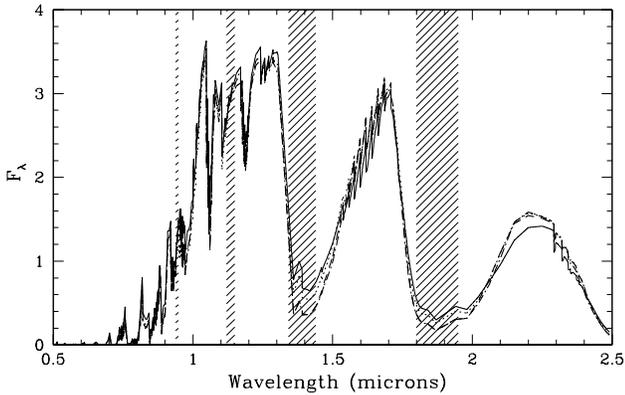}
\caption{In this figure we have plotted four models from
different
series : O10 - solid; E08 - dotted; M05 - short dash; P05 -
dot-short dash. The shaded portions define the wavelength
zones which are strongly affected by telluric absorptions and
hence are not included while comparing the spectra.}
\label{simmods.fig}
\end{figure}

Noticeable features in the synthetic spectra are the varying 
shapes of the H$_2$O bands. This is illustrated 
in Figs.\,\ref{shape.fig} and \ref{simmods.fig}.
In the top panel of Fig.\,\ref{shape.fig}, 
the near-maximum P models  are plotted, and in the lower panel the
near-minimum models. The H$_2$O bands centered around 1.4 and 1.9\,$\mu$ms 
are not only deeper but also wider for the near-minimum
models. In their discussion on water `shells', TLS03
show that for these near-minimum models the shells are located relatively
close to the continuum forming layers and the disk brightness
distributions show protrusion-type (i.e. stronger H$_{2}$O)
two-component features. The top panel of Fig.\,\ref{shape.fig}
and Fig.\,\ref{simmods.fig} illustrates that models with otherwise
similar colours may have differing water band depths or shapes,
as expected for complex atmospheric stratifications. 
In some models, H$_2$O appears
in emission in near-IR portions of the spectrum.
This means that fluxes in water-covered regions of the spectrum 
are higher than computed pure-continuum fluxes (see Fig.\,\ref{apend.fig}
in the Appendix). When of moderate strengths, such emissions 
will not be readily recognizable in observed low-resolution spectra. 
Since water is treated as a strict-LTE absorber and temperatures
decrease monotonically with radius, these emissions are a mere
large-volume effect resulting from a favourable combination of the
geometry, the optical thickness and the temperature of the
atmospheric water `shell' (cf. TLS03).
In a few near-maximum P models (e.g. P18, P20, P38), we
see a more easily identifiable enhanced emission component 
in the wings of the water bands. A semi-empirical study of observed 
emissions in Miras between 3.5 and 4.0\,$\mu$m was given by 
Yamamura et al. (1999) and Matsuura et al. (2002).

Fig.\,\ref{simmods.fig} also illustrates the degeneracy between
model parameters in the determination of the output spectra: in
this example, similar
spectra are obtained although the model masses vary between 1 and 
2\,$M_{\odot}$, luminosities between 1650 and 7070\,$L_{\odot}$,
radii between 290 and 510\,$R_{\odot}$, and phases between 0.5
and 1.0. Only the effective temperature $T_{\rm 1.04}$ is relatively 
constant in this specific case, with extremes separated by $\sim$ 200\,K.
Degeneracies such as the one shown are common among the models. 
Details of the model spectra have to be given a high level
of confidence if one wishes to discriminate further between the 
various parameter sets. To set a confidence limit
one must consider the quality of the best fits between models and spectral data,
as done in the two following sections. It then becomes
clear that differences such as those in Fig.\,\ref{simmods.fig} 
currently are too small to be assessed reliably: the exercise of
estimating fundamental parameters from a spectrum will not yield
unique solutions.  

%The uncertainties in the 
%observed spectra and those assigned to the model predictions
%determine whether or not two predicted spectra
%are significantly different. 
%With the spectra currently available,
%the exercise of estimating fundamental parameters does not
%yield unique solutions. 
 
%------------------------------------------------------------------
\subsection{Observed Spectra}
\label{observed.sec}
The sample of observed spectra of oxygen-rich Miras is taken
from LW2000. We refer the reader to this paper for details on
observation and data reduction. The parameters of the observed
Miras are listed in Table 3 of LW2000. They span a range of
pulsational properties and estimated masses and metallicities.
From this sample we have selected all Mira spectra covering 
the entire range from 0.5 to 2.5\,$\mu$m.
For many of the Miras there are several such spectra, taken
at different phases, which gives us a larger database (48 complete
spectra for 16 O-rich Miras). The high
frequency noise in the spectra is low, with a typical 
signal-to-noise ratio per resolved element of 50 (except
in regions of heaviest telluric absorption). For colours
combining $K$ with a magnitude measured around 1\,$\mu$m,
the 2$\sigma$ uncertainties are about 0.2 magnitudes, and those
on $(V-K)$ amount about 0.3 magnitudes. The phases of observations 
are uncertain for many sources. Amateur light curves
from AAVSO\footnote{American Association of Variable Star Observers} and 
AFOEV\footnote{Association Fran\c{c}aise des Observateurs d'Etoiles Variables} 
were used when available (cf. LW2000). 
In other cases, phase could be grossly re-evaluated from the available
spectra themselves, using the general anti-correlation between
luminosity and colour-temperature and the occurrence of 
Paschen\,$\beta$ emission near maximum light.
%----------------------------------------------------------------
\section{Comparison with observed data}
\label{compare.sec}
In this section, we compare the global shape of the observed and
the model spectra. For the purpose of a detailed discussion,
we restrict ourselves to three observed Miras that have similar 
periods to those of the models, namely R Cha ($P\simeq 334$\,d,
$\Delta M_{\rm V}\simeq 6.7$), RS Hya ($P\simeq 338$\,d, $\Delta M_{\rm
V}\simeq 5.2$) and CM Car ($P\simeq 338$\,d, $\Delta M_{\rm V}\simeq 5.2$).
The foreground extinction for these stars was estimated from the
reddening maps of Burstein \& Heiles (1982). The authors 
quote an uncertainty of $\sim 10\%$ on the map values. In the
absence of {\em Hipparcos} parallax measurements, 
we estimate the distances to these three Miras from the ($K, \log P$) 
relation of Hughes \& Wood (1990). The derived values of $A_{V}$ are $\sim$ 0.4,
0.2 and 0.6 respectively. A conservative error estimate 
of a factor of 2 on distance and the given uncertainty in
extinction measurements would modify $A_V$ by 0.1\,mag in the worst case,
all three stars being high enough above the galactic plane.
The dereddened spectra are used for the comparative study discussed below.

\begin{figure*}
\includegraphics[width=1.0\textwidth,height=0.95\textwidth]{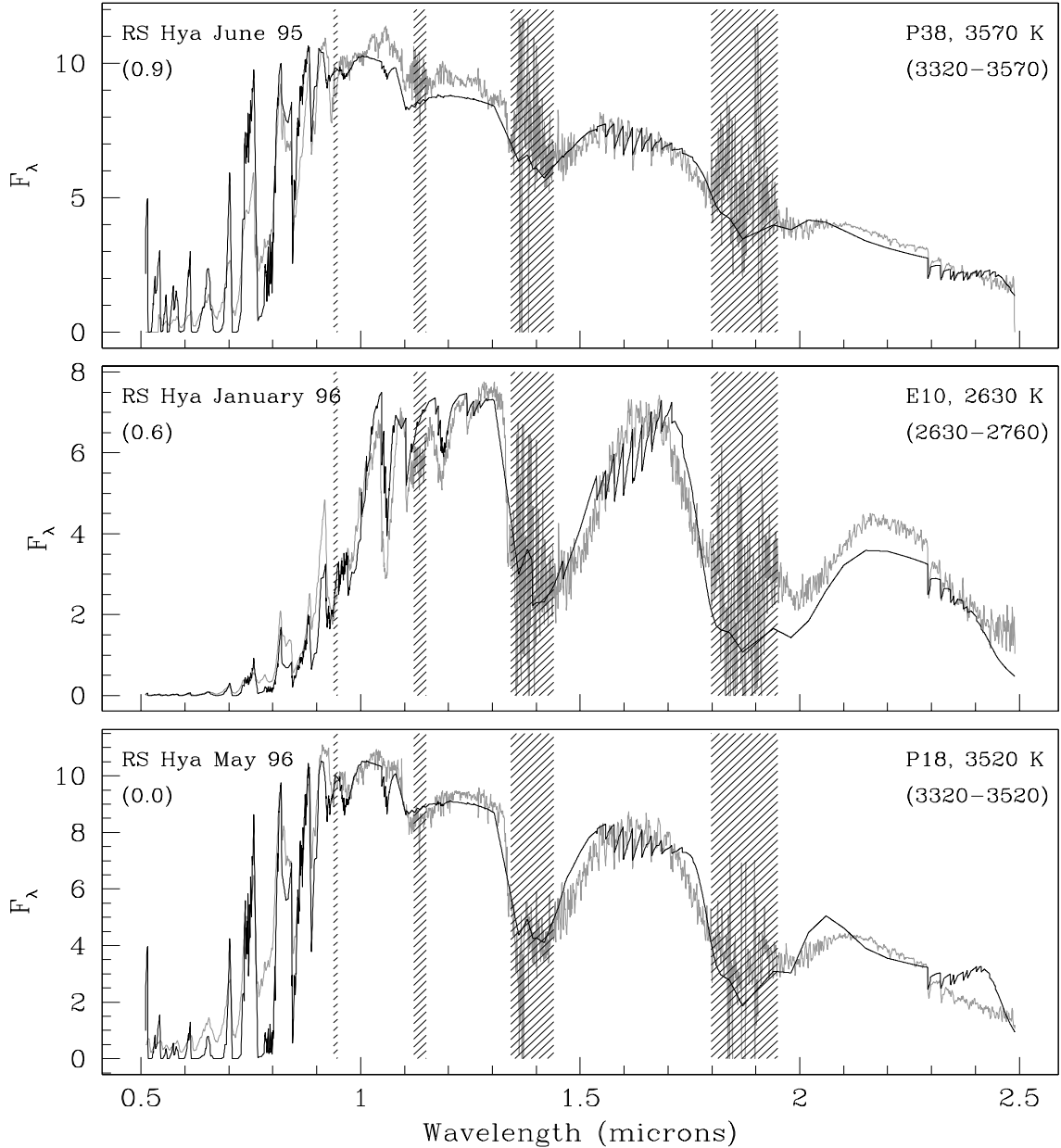}
\caption{This figure shows the empirical spectrum (grey) 
and one of the best fitting models (black) for RS Hya 
as observed in June 1995 (upper panel), January 1996 (middle panel) and 
May 1996 (lower panel).
The flux densities per unit wavelength are in arbitrary units.
The estimated phase of observation is given on the left side of
each plot.  On the right side, the plotted model is identified 
(see Table\,\ref{param.tab}). The range of temperatures ($T_{1.04}$)
reached by models that provide fits of similar quality is indicated underneath.
Shaded strips are the regions of strong telluric absorption which are 
excluded during comparison.}
\label{comp1.fig}
\end{figure*}
\begin{figure*}
\includegraphics[width=1.0\textwidth,height=0.95\textwidth]{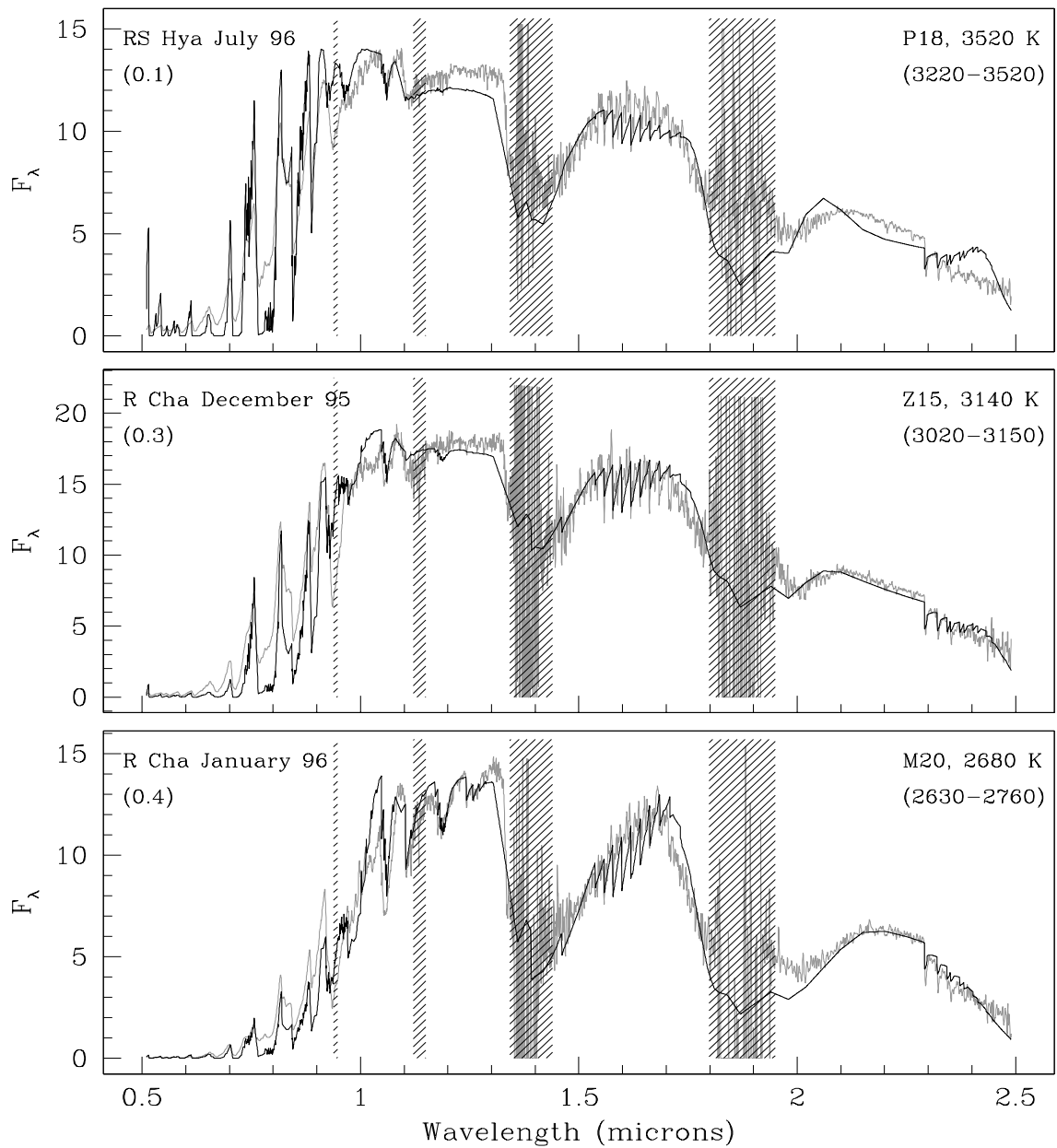}
\caption{Same as Fig.\,\ref{comp1.fig} but for RS Hya in July 1996 
(upper panel), and for R Cha in December 1995 (middle panel) and 
January 1996 (lower panel).}
\label{comp2.fig}
\end{figure*}
\begin{figure*}
\includegraphics[width=1.0\textwidth,height=0.95\textwidth]{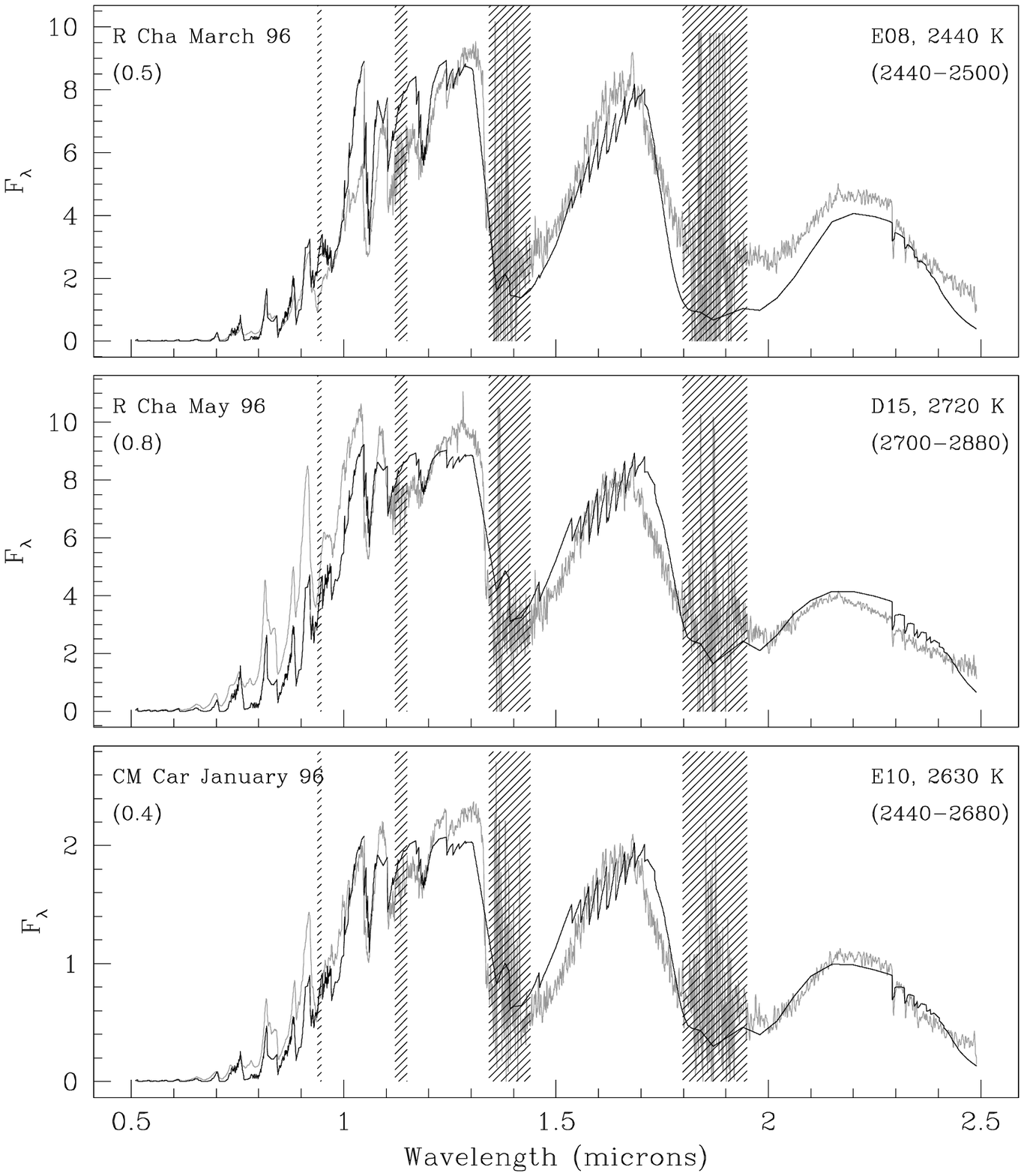}
\caption{Same as Fig.\,\ref{comp1.fig} but for R Cha in March 1996 
(upper panel) and  May 1996 (middle panel), and for CM Car in January
1996 (lower panel). Note that for the March 96 spectrum of R Cha the 
$T_{\rm eff}$ values of the `best fit' models reach as low as 2160\,K
($T_{\rm 1.04}$ values are given in the figures).} 
\label{comp3.fig}
\end{figure*}

For a first, automatic selection of appropriate models, 
we compare each observed spectrum with the entire set of
model spectra and quantify the quality of each fit with
a $\chi^{2}$ value. In computing $\chi^{2}$, care is taken to
exclude portions of the spectra which are strongly affected by
telluric absorption (see Fig.\,\ref{simmods.fig}). 
Otherwise,
equal weight is given to each data pixel (i.e. 950 points below 1\,$\mu$m
and 2400 points above). Rather than rejecting all models but
the one with the minimum $\chi^2$, we keep the 5 to 10 models 
with $\chi^2$ values not significantly larger than the minimum
value. This is justified as follows. 
Mathematically speaking none of the resulting
$\chi^2$ values is good\,: the difference
between the data and the best fitting models cannot be 
explained by observational errors alone. In that situation, the search for 
best fits still makes sense, but the definition of the rejection 
criterion becomes more arbitrary. Different `best' solutions can be obtained
by giving different features (band strengths, band shapes, continuum
colours, etc.) different relative weights in the $\chi^2$. 
Our threshold on the naturally-weighted $\chi^2$ value was set so that 
the selected models approximatively cover the range one would obtain by 
modifying weights in various reasonable ways. 

Eye inspection followed the automated pre-selection, taking into
account the following arguments.
The model reproduction of the spectral features are
considerably better in the near-IR as compared to the region
below 1\,$\mu$m. There are various conceivable model inadequacies
in the optical domain responsible for this disparity. 
First, the simple treatment of TiO bands dominating
the optical spectrum may result in appreciable inaccuracies of
band depths, in particular for very strong features (Section 2.1;
cf. BSW96). Second, the pronounced temperature contrast between
continuum-forming layers and the upper layers where the molecular
bands are formed makes the shapes and depth of these TiO bands
strongly dependent on the subtle details of the outer atmospheric
stratification in the Wien part of the spectrum. Third, if dust
particles were formed in the outer atmosphere, this would
strongly affect the optical spectral region (Bedding et al. 2001;
cf. Section 4.2).
%These inadequacies are kept in mind while visually selecting
%the few `best fitting models'. Low weight is given
As a consequence of the above, low weight is given to the quality
of fits to the individual spectral features in the optical part
of the spectrum. However, it should be noted that the overall
energy distribution in the region below 1\,$\mu$m and the 
ratio of optical to near-IR light do play a
significant role in constraining the model fits.

As the result of the above selection, we are left for each observed
spectrum with a few model spectra that qualify as satisfactory\,:
the energy distribution, the relative band strengths and shapes 
are reproduced to a level that was unexpected considering the
complexity of these objects and the limited range of explored
model parameters. Because of the degeneracies in the models
mentioned previously, the qualifying models span a wide range of 
parameters. Even the effective temperature shows significant dispersion
among acceptable solutions.
%We therefore have to refrain from presenting
%actual parameter values in this paper.

In Fig.\,\ref{comp1.fig}, \ref{comp2.fig} and \ref{comp3.fig}
we show comparisons with typical `best fitting' model spectra. 
For RS Hya, the four spectra shown are reproduced reasonably well by the
models. The best fitting models for the January 1996
spectrum deviate beyond 2\,$\mu$m. These
deviations in the energy distribution are consistent with the estimated 
observational uncertainties at the $2\sigma$ level only.
In the case of R Cha, the agreement between 
the models and the observed spectra is very good for the
December 1995, January 1996 and May 1996 spectra.
The March 1996 spectrum of this star stands out with a particularly
red slope around 1 -- 1.3\,$\mu$m and a corresponding
excess beyond 2\,$\mu$m, a behaviour that resembles the January 1996
spectrum of RS Hya. It is also striking that these two spectra
have particularly red slopes between 1 and 1.3\,$\mu$m. We discuss
them in detail in Sect.\,\ref{1mu.sec}. For CM Car, the best
fits are also satisfactory.

We re-evaluated the visual phases of the observed stars based on
the appearance of the available time series of observed
spectra. A phase diagnostic feature is the
Paschen\,$\beta$ line which appears in emission near maximum (Fox et al. 1984).
The re-estimated phases are uncertain to $\pm$0.1 cycles.
For the CM Car spectrum the uncertainties could be larger as
there is no time series available.  For the models, the visual
phase was estimated from the bolometric light curves 
by BSW96 and HSW98, assuming an offset of
about 0.1 cycles between the two (Lockwood \& Wing
1971). Uncertainties of this offset estimate, 
irregular shapes of luminosity curves (see Figs. 1 to 3 of BSW96 and HSW98), 
and deviations of up to several days between the formal pulsation period of the 
parent star and actual mean periods of pulsation series, may 
readily result in phase uncertainties of the order of 0.1 to 0.2.

The phase matching between the observed spectra and the
best-fitting model spectra shows random discrepancies. 
For most cases, the agreement between the observed phase and the
phase of the best-fitting model spectra is within 0.3 cycles.
This is consistent with the two quoted uncertainties.
There are exceptions however (e.g. the December 1995 spectrum of R Cha),
where the set of good model fits covers a wide range of phases. This
is not surprising in view of the degeneracies illustrated in 
Fig.\,\ref{simmods.fig}. 

We carried out this model fitting procedure for the other Miras
in the sample, which have different periods, and found similar 
levels of agreement for more than 80\,\% of the observed spectra.
In summary, the overall shape and the gross features of the observed 
spectra are well reproduced by the models. Most broad band 
energy distributions can be adjusted to within the observational
uncertainties, despite the limited parameter and phase or
cycle coverage of the available models (see also Sect.\,\ref{bbc.sec}). 
The shapes of wide molecular bands can be adjusted very well
when the comparison is restricted to a subsection of the spectrum.
The constraint set by the complete energy distribution is heavy,
and resulting imperfections of the best fits must be blamed 
at least partly on the small size of the model set.
%In the near-IR, the models are at a low resolution compared to the
%observed spectra which restricts us from investigating finer features. 

Considering the models and data as samples and looking at
specific spectral features will throw more light on the
systematics in the respective behaviours of the models and the observations. 
This is the aim of the following section.
 
%------------------------------------------------------------------
\section{Behaviour of specific spectral features}
\label{features.sec}
It is beyond the scope of this paper to study the numerous
features of Mira spectra individually. We have
chosen to focus on three specific features: (1) the broad band
colours, because they are the most commonly measured quantities;
(2) the spectral region between 1 and 1.3\,$\mu$m, which apart
from displaying various slopes both for the models and the
observed spectra is currently considered to contain
near-continuum wavelength regions; (3) a few near-IR molecular
bands for which specific narrow band filters exist, thus making
them more accessible to observations. Among the latter, H$_{2}$O is given
most attention because of its potential relevance to the
interpretation of the near-IR angular diameter measurements
(TLS03).

\subsection{The broad-band colours}
\label{bbc.sec}
\begin{figure*}
\includegraphics[width=1.0\textwidth]{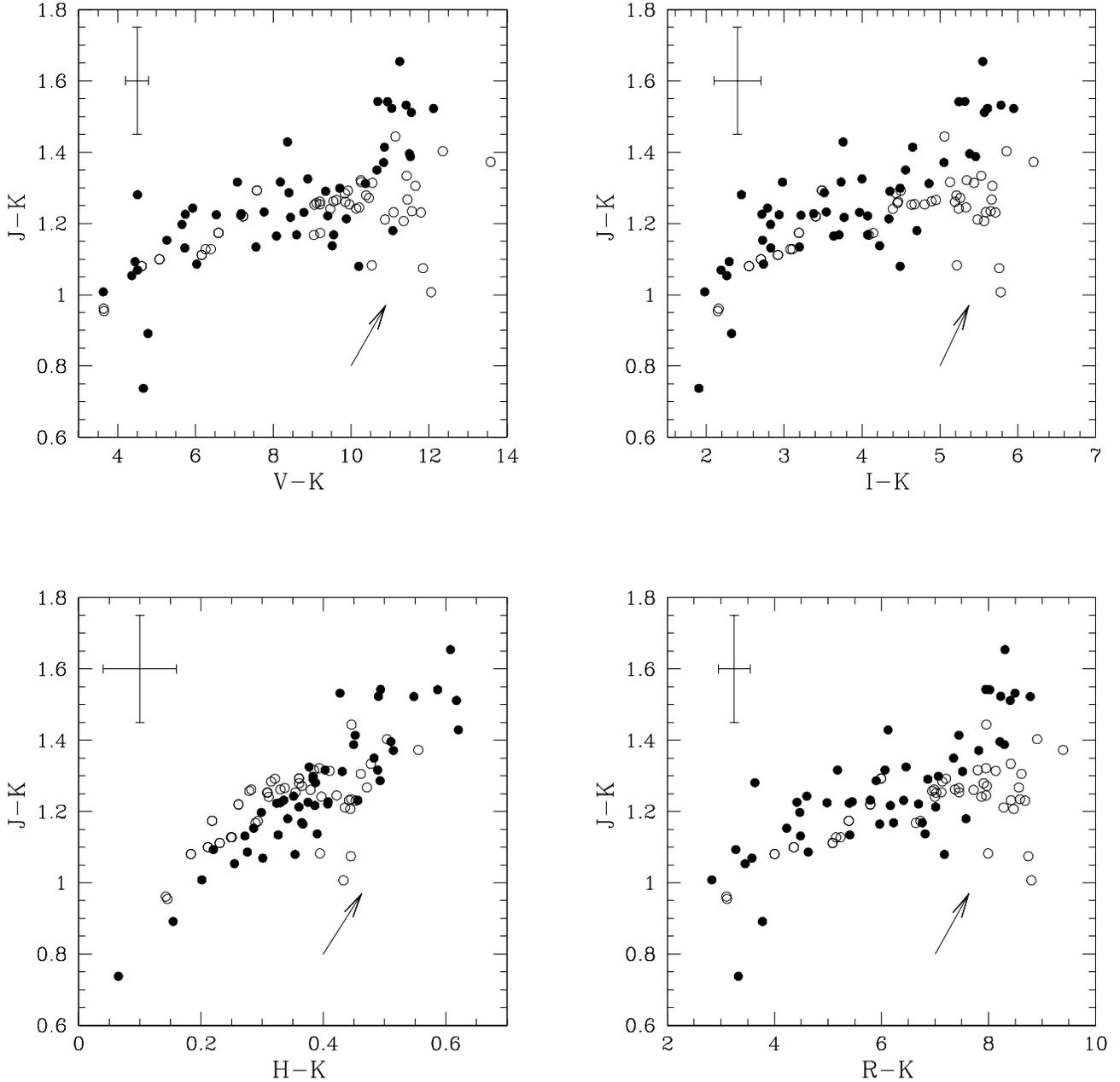}
\caption{Broad-band colours for the
observed spectra (solid circles) and the model spectra (open circles). 
Also shown are the 2$\sigma$ uncertainties of the
observed colours and the Milky Way extinction vector for A$_{V}$=1.}
\label{bbc.fig}
\end{figure*}

For cool stars the observed broad-band colours and magnitudes are
often used for estimating luminosities and effective temperatures.
In this section, we investigate the trends seen in the observed
and model predicted colours. Fig.\,\ref{bbc.fig} shows 
a set of broad-band colours derived from the model and the
observed spectra. No extinction correction is applied to the
colours on this figure. The arrows indicate the effect of interstellar 
extinction for A$_{V}$=1 (greater than the values derived for the three
stars of Figs.\,\ref{comp1.fig}, \ref{comp2.fig} and \ref{comp3.fig}). 
Also shown are the 2$\sigma$ uncertainties of the 
observed colours. Given the scatter, the interstellar
extinction would not appreciably
affect the trends seen. Within these uncertainties, the plots 
show considerable overlap between the loci of the observed and 
the model predicted colours. Systematic offsets between the colours 
predicted by the models and those observed, if any, are small.

Despite the limited range of explored model parameters and the
uncertainties involved in modeling the Wien part of the
spectrum, the extent of observed broad-band colour distributions is reproduced
rather well. The largest difference is seen in $J-K$ and
$H-K$, where the observations reach redder colours than the models.
The selection of spectra with $J-K>1.45$ leads to a homogeneous subsample 
of spectra similar to the March 1996 observation of R\,Cha
and the January 1996 observation of RS\,Hya, two spectra 
already pointed out in Sect.\,\ref{compare.sec}. These two spectra show a
red slope in the 1\,$\mu$m region and also display IR excess beyond
2\,$\mu$m when compared with dust-free model spectra. Anticipating
the discussion of Sect.\,\ref{1mu.sec}, we note that this hints
at the presence of reddening, possibly due to the presence of atmospheric
and circumstellar dust, or at the existence of additional sources
of atmospheric opacity that may be currently missed or
underestimated in the models.

%----------------------------------------------------------------------------
\subsection{The spectral slope between 1 and 1.3\,$\mu$m}
\label{1mu.sec}
A close study of the nature of the model and observed
spectra reveals distinct differences in the slope of the 1\,$\mu$m 
region. The models conspicuously display rather
flat spectra in this region compared to the variety of slopes found
in the data. The spectral window around 1.04\,$\mu$m 
is a particularly important part of the spectrum of Mira-type stars, as it is
often considered one of the best access points to the 
near-IR continuum (Jacob \& Scholz 2002). The nature and possible
origin of this disparity in slope is the focus of this section.

Blue slopes in the region between 1 and 1.3\,$\mu$m were commented
upon by LW2000. While studying phase offsets between the cycles
followed by the strength of VO (1.05\,$\mu$m) and other spectral signatures
(cf. Alvarez \& Plez 1998), they noted that the blue slopes 
tend to occur just before the disappearance (or strong reduction) of VO
and other molecular bands. Although their argument was based on 
a small number of stars for which phase coverage was sufficient,
they suggested that this happens just before, or in any case close to 
maximum light (spectra that still display Paschen lines 
but show no remaining VO absorption occur slightly later 
in the same pulsation cycle).

The new phase 0.8 models computed for the P series indeed
display blue slopes in the 1\,$\mu$m region, 
while still displaying deep molecular
bands characteristic of low temperatures (see Appendix A). As
shown in Sect.\,\ref{compare.sec}, a significant number of observed
spectra can only be adjusted with models having these properties
(Figs.\,\ref{comp1.fig}, \ref{comp2.fig} and \ref{comp3.fig}).
The 0.8 models of the Z series also display blue $1-1.3\,\mu$m slopes but 
this is less surprising as it agrees with an otherwise warm energy distribution.
In the models, these blue $1-1.3\,\mu$m slopes appear only shortly before
maximum light when the newly emerging shock front is still 
in deep layers and one can probe the deep hot layers. 
This result clarifies the empirical trend found by LW2000.
%Phase matching of the observed and the model spectra
%shows that such blue slopes are indeed seen close to
%maximum, thus confirming the suspicion of LW2000.

\begin{figure}
\includegraphics[clip=,width=0.5\textwidth]{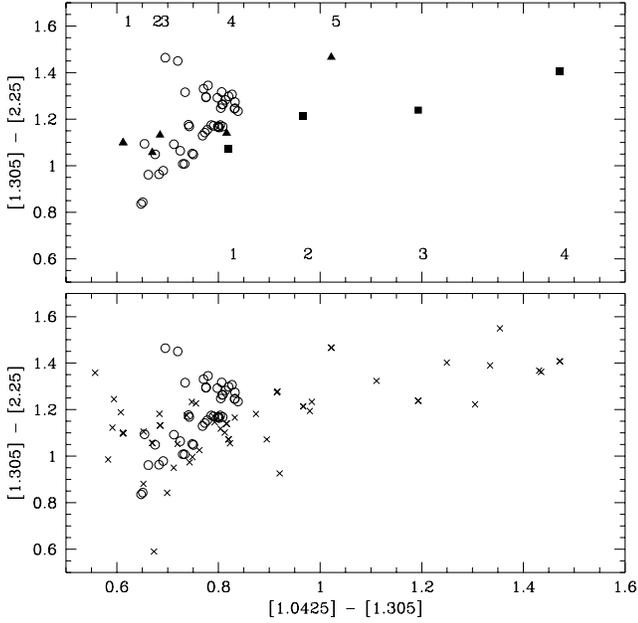}
\caption{Lower panel: Model predicted (open circles) and observed
(crosses) narrow-band colours. [1.0425] -- [1.305] measures the slope
in the $1-1.3\,\mu$m region, [1.305] -- [2.25] is a narrow-band analog 
of J-K.  Upper panel:  Here, the models (open circles)
and two individual Miras RS Hya (solid triangles) and R Cha (solid squares)
are plotted. The numbers at the top indicate the phase of
observation of the RS Hya spectra; 
(1:2:3:4:5 correspond to 0.8:0.9:0.0:0.1:0.6) and the numbers on 
the lower side are for R Cha; (1:2:3:4 correspond to 0.8:0.3:0.4:0.5). }
% old, wrong: 0.4:0.5:0.8:0.3). }
\label{1mu_mod_data.fig}
\end{figure}

The global behaviour of the slope in the 
$1-1.3\,\mu$m region is illustrated
in the narrow-band [1.0425] -- [1.305] versus [1.305] -- [2.25]
colour-colour plot shown in the lower panel of 
Fig.\,\ref{1mu_mod_data.fig}. The colours are measured through
square filters centered at 1.0425, 1.305 and 2.25\,$\mu$m,
with respective full widths of 5, 20 and 20\,nm\,; they are 
given in magnitudes and take the value 0 for Vega.

The dynamical models (open circles) occupy a very narrow
range of values in [1.0425] -- [1.305] compared to the observed data
(crosses). The dynamical ranges of the [1.305] -- [2.25] colours
are reasonably consistent for the model and observed spectra.
The observed spectra that show the bluest [1.0425] -- [1.305]
colours in the plot (i.e. bluest slope in the 1 -- 1.3\,$\mu$m
region), also display Paschen line emission
and deep molecular bands. Spectra with
warm energy distribution and shallow molecular bands have small
values in both the pseudo-continuum colours and 
hence occupy the lower left quadrant of the plot.

The lower panel of Fig.\,\ref{1mu_mod_data.fig} also shows that
several observed spectra have very red slopes in the $1-1.3\,\mu$m 
region, while there are no models with [1.0425] -- [1.305] 
greater than 0.9. The observed spectra with such red slopes
include and resemble the January 1996 spectrum of RS Hya and
March 1996 spectrum of R Cha (cf. Sect.\,\ref{compare.sec}). 
The upper panel of Fig.\,\ref{1mu_mod_data.fig} gives a comparative
view of the multi-epoch spectra of these two selected Miras and
the model spectra. 
%The numbers displayed in the upper and lower
%part of the plot are indicative of the phases of observations of
%RS Hya and R Cha respectively and are explained in detail in the
%figure caption. 
From this plot it is clear that the variation 
of the colour [1.0425] -- [1.305], which is indicative of the 1\,$\mu$m
slope, with phase is not only larger for the observed Miras as
compared to the dynamical models, but also the variation is along
a different direction. 
\begin{figure}
\includegraphics[clip=,width=0.5\textwidth]{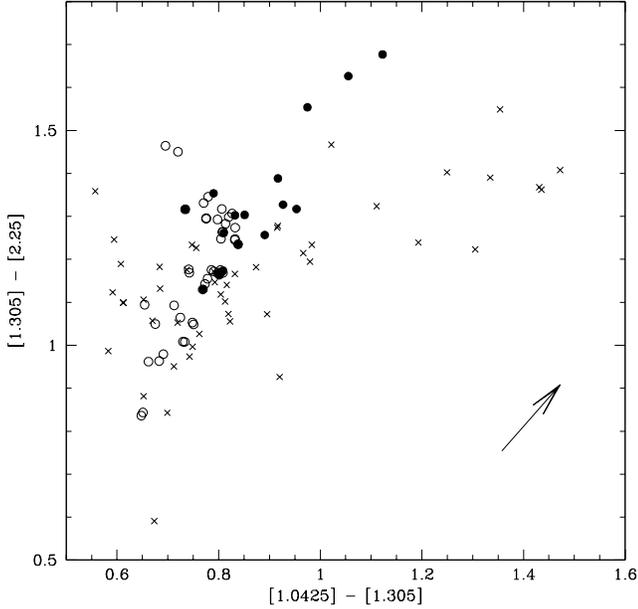}
\caption{Location of the dusty models (solid circles) in the two-colour 
plot of Fig.\,\ref{1mu_mod_data.fig}. Open circles are dust-free models 
and crosses are observations. The arrow corresponds to $A_v=1$ with 
the Milky Way interstellar extinction law.
}
\label{1mu_mod_dust.fig}
\end{figure}

Opacity data in the 1\,$\mu$m region might be the most likely
origin for the lack of red model slopes. The current models may
overlook or underestimate sources of opacity.
Independent evidence in favour of additional opacities at large
radii comes from wavelength and phase-dependent angular diameter
measurements (e.g. Young et al. 2000 for $\chi$ Cygni).  
The sources of such opacities could be molecules or dust-grains
which would significantly affect both the spectra and the
disk brightness distribution.

The molecular sources of opacities of the models were assembled
according to their importance in normal red giants, and may
indeed be incomplete at low temperatures reached in outer Mira
atmospheres. Relevant molecular absorption sources that are
already included, but for which the adopted opacities may be
incomplete, are TiO, VO and H$_{2}$O. The choice of one or the other 
particular combination of the various published
opacity tables for these three
molecules plays a decisive role in cool dwarf star models (P.
Hauschildt, private communication). One may also extend the list
of potential molecular opacity sources to more exotic species
such as YO, ZrO or ScO, that reach observable proportions in S
star models (e.g. Piccirillo 1980), or to molecules observed in
dwarfs or subdwarfs (CN, HCN, CaH, FeH to cite a few; see e.g.
Allard \& Hauschildt, 1995, for a more extended list). 
We have not explored these possibilities, hence
it is beyond the scope of this paper to give a quantitative
discussion on the molecular opacity issue and this point
is left open to future studies. 

\begin{figure}
\includegraphics[clip=,width=0.5\textwidth]{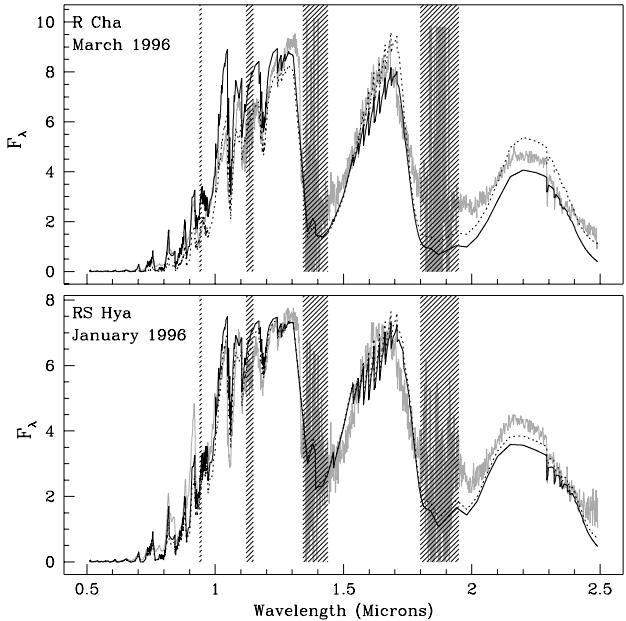}
\caption{This figure shows the March 96 spectrum (grey) of R Cha (upper
panel) and the January 96 spectrum (grey) of RS Hya (lower panel). Also plotted
are the best fitted dust-free (solid) and dusty (dotted) models. Fluxes are 
given in arbitrary units of energy per unit wavelength.}
\label{rcha_rshya_dust.fig}
\end{figure}

The formation of dust in the outer atmospheres of at least some
Miras, on the other hand, may be considered a fact supported by
the evolutionary transition from Miras to dust-enshrouded OH/IR
sources. Ferguson et al. (2001) give reasons for dust formation
and absorption around cool static red giants. Observationally,
some interferometric and spectroscopic measurements indicate that
inner edges of dust shells might be as close  2 or 3 continuum
radii from the star's centre (e.g. Danchi et al. 1994; Danchi \&
Bester 1995; Lobel et al. 2000; Lorenz-Martins \& Pompeia 2000).
Bedding et al. (2001) have studied the occurrence of dust in the 
atmospheres of Miras and its effect on their spectra. 
Dust thermodynamically fully coupled to 
the surrounding gas is shown to affect the 1.04\,$\mu$m passband 
and to distort the brightness distribution considerably. 
This prompted us to investigate the role of dust in the 1\,$\mu$m
region. In Fig.\,\ref{1mu_mod_dust.fig}, we show the loci of
the dusty models from Bedding et al. (2001) in the same two
colour diagram as in Fig.\,\ref{1mu_mod_data.fig}. 
The blue end of the dusty models overlaps with the
dust-free models. The dusty models do show redder slopes and display
large [1.0425] -- [1.305] values but they also have a strong excess in the
[1.305] -- [2.25] colour. Nevertheless, we compared the spectra of the
three Miras presented in Figs.\,\ref{comp1.fig},
\ref{comp2.fig} and \ref{comp3.fig} with the dusty models. Only
for two spectra, namely the March 1996 spectrum of R Cha and January 1996
spectrum of RS Hya better agreement is obtained with the dusty models.
These are the spectra already identified as particularly red
in Sect.\,3, and they are also representative of all the 
spectra with particularly large values of $J-K$ and
$H-K$ in Fig.\,\ref{bbc.fig}. The two new `best fits' 
are shown in Fig.\,\ref{rcha_rshya_dust.fig}.

In view of the above comparison of observed spectra with the dusty models,
a few important points are worth mentioning.
Firstly, Bedding et al. (2001) is an exploratory study on 
the occurrence of dust only in a small subset of the available
models, and they have tested only a few values of the dust 
condensation fraction. Therefore remaining imperfections in the fits 
of the energy distributions of the two stars of Fig.\,\ref{rcha_rshya_dust.fig}
are not surprising. The changes induced by the presence of atmospheric
dust go in the right direction. 
Bedding et al. (2001) have also adopted a particular dust composition\,;
other assumptions would lead to somewhat different effects on
the colours (the arrow in Fig.\,\ref{1mu_mod_dust.fig} shows the effect of
average Milky Way interstellar dust). Secondly,
the fact that for most of the observed spectra the fit quality 
is not improved with the few available dusty models does not by itself exclude
the presence of some atmospheric or circumstellar dust.

There is some indication of variation of dust
with phase but it is difficult to make any quantitative statement
regarding this in the absence of proper phase coverage of
the dusty models. This can be better appreciated by looking at the
loci of the two individual stars in Fig.\,\ref{1mu_mod_data.fig}.
For R Cha and RS Hya the dusty models fit better
to phases 0.5 ($\pm$0.1) and 0.6 ($\pm$0.1) respectively. 
For R Cha the rest of the
observed spectra (phases 0.3, 0.4 and 0.8) have similar level of
fits with dusty and dust-free models. For RS Hya dusty models do not 
provide good fits to the rest of the observed spectra (phases 0.8, 0.1, 0.9
and 0.0). This is consistent with a general scenario in which dust effects
are maximum at minimum light and hence at the lowest temperatures. 
To conclude the above discussion, we consider it likely that the 
disparity seen in the slope around 1\,$\mu$m between the models 
and the observed spectra is partly due to dust.
%---------------------------------------------------------------------
\subsection{The molecular bands}
\label{mol_bands.sec}

\begin{figure*}
\centerline{\includegraphics[width=0.95\textwidth]{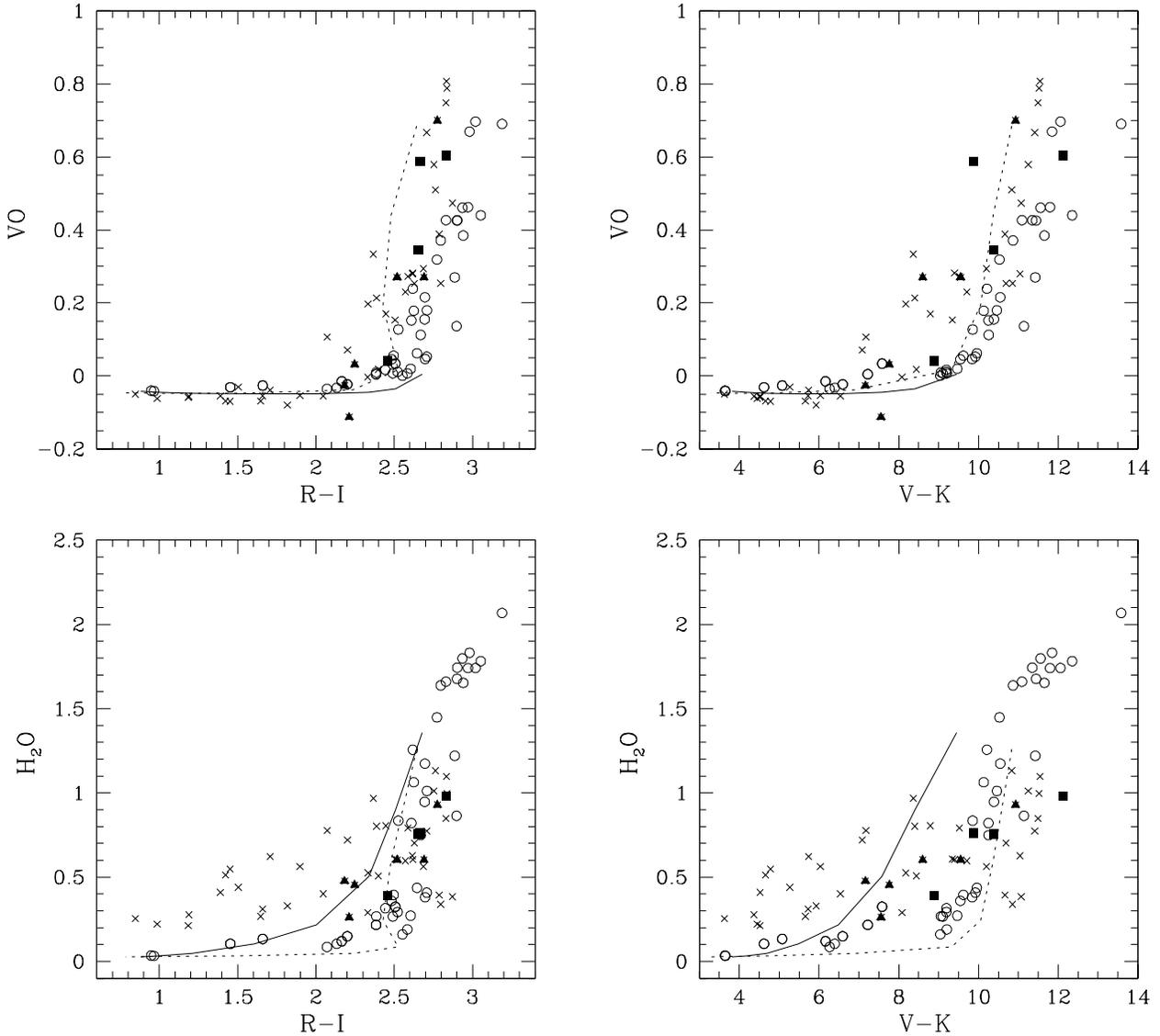}}
\caption{H$_{2}$O and the VO indices 
as a function of the $R-I$ and $V-K$ broad-band
colours for the model (open circles) and the observed (crosses)
spectra. The data points for RS Hya and R Cha
are highlighted, respectively, 
as solid triangles and squares. Overplotted are the static
giant models XX (metallicity +0.5: dotted line) and YY
(metallicity -1.0: solid line).}
\label{h2o_vo.fig}
\end{figure*}

The spectrophotometric features due to H$_{2}$O and VO are amongst 
the most prominent molecular bands seen in the spectra of Miras. 
In this section, we compare these bands in
the observed and the model spectra. For band measurements,
we use narrow-band filters similar to those of Wing (1967) and 
White \& Wing (1978), adopting the
passbands defined by Bessell et al. (1989; see Fig.\,3 of Alvarez et
al. 2000). The indices are the ratios
between the fluxes measured through a passband in the molecular band
and a passband that is not or little affected; like colours, they are expressed
in magnitudes and take the value 0 for Vega. The VO band at 1.05\,$\mu$m is
measured by the [105] -- [104] colour index. For H$_{2}$O,
we consider the index [220] -- [200] located in the wings of the 1.9
\,$\mu$m band. In Fig.\,\ref{h2o_vo.fig},
we plot the two molecular indices as a function of two
broad-band colours. 

The values of the VO index span similar 
ranges for the models and the observations, which is satisfactory. 
A small but significant offset in colour between models and data
is present as one moves to the red ends of the distributions. 
The value of this offset is larger in the plot of VO {\em vs.} $V-K$, as
expected from the relation between $V-K$ and $R-I$ 
(which agrees well between model and data for the large values of 
$V-K$ that are relevant here).  At a given colour, observed VO bands
appear stronger than the model predicted ones. 
The Wing VO index considered above measures the molecular
absorption at 1.054\,$\mu$m, i.e. somewhat off the band's
deepest point (near 1.06\,$\mu$m in the models). If we use a modified index 
with a more centrally located passband, we find that the VO values for the
models increase by about 0.1 magnitudes in the VO {\em vs.} colour plots,
while the values for the empirical spectra remain essentially unchanged.
The offset is slightly reduced. The fact that data and models have 
different sensitivity to a change in the adopted index is due to the
shape of the modeled absorption bands, which do not precisely agree
with the observed shapes. Here, the opacity data (and possibly the
overlap with bands from various molecules) are the most likely origin 
of the discrepancies.

For the H$_{2}$O index we note that the 
overlap between the observed and the model points is
less satisfactory. The plot shows that for blue $R-I$
and $V-K$ colours the observed spectra display stronger H$_{2}$O bands 
compared to the model predicted values (see Matsuura et al. 1999 for
additional observations of early M type stars with significant water bands).
On the other hand the models show strong H$_{2}$O bands 
($>$ 1.3) towards the red end and no observations are found in that region
of the diagrams.
The models with these strong H$_{2}$O bands are the near-minimum P and M
models and all the O models.
The range of model-predicted H$_{2}$O indices is much larger than
the variation seen with phase for individual observed Miras. 

Random selection
of an observed spectrum and a model spectrum with similar H$_{2}$O index,
shows good agreement in the shape and strength of the H$_{2}$O bands at
1.4\,$\mu$m and 1.9\,$\mu$m. Different indices, designed to measure
absorption in the shorter
wavelength wing of the 1.9\,$\mu$m band and the wing of the 1.4\,$\mu$m
band, show similar trends in the H$_{2}$O {\em vs.} colour plots. These
two arguments indicate that the disparity seen in the plots is not due to
the choice of one particular index. 

The other parameters in the plots of Fig.\,\ref{h2o_vo.fig}
are the broad-band colours.
Fig.\,\ref{bbc.fig} doesn't reveal any obvious problem with
the colours.  The uncertainty estimates on the observed broad band 
colours by LW2000 have been checked and found correct. However,
in addition to these random errors there might be systematic
errors in the merging procedure between optical and near-IR data,
in particular for spectra with very steep energy distributions
near 1\,$\mu$m (see models of Fig.\,\ref{shape.fig} or \ref{simmods.fig}).
This might have led to systematically bluer $V-K$ values
for these particular spectra.  Going back to the original data, we estimate
that 0.5\,mag is an upper limit of such errors, if present
(such an effect is not incompatible with Fig.\,\ref{bbc.fig}). For stars
without such dramatic features around 1\,$\mu$m, we find no reason for 
systematic errors. The merging procedure would not affect
the $R-I$ colour. We therefore conclude that observational 
uncertainties in the broad band colours might contribute but
cannot by themselves explain the offsets in the index-colour plots.

Metallicity could be one of the reasons for the disparity in the
plot of H$_2$O {\em vs.} $V-K$.
The pulsating star models are computed only for solar metallicity. 
To show the likely effect of metallicity, we have plotted lines 
showing static giant models with two extreme metallicites
(Z/Z$_{\odot}$=+0.5 and -1.0) but identical masses and luminosities
(1\,M$_{\odot}$, $10^4$\,L$_{\odot}$; Bessell et al. 1989, 1991;
see Bessell et al. 1991 for loci of evolutionary tracks). 
The lower metallicity models have warmer colour temperatures.
In the dynamical case, it is likely that lower metallicity also 
shifts the locus of the dynamical spectra to bluer colours, while pulsation
might still allow for strong H$_{2}$O bands.
A negative correlation between the H$_{2}$O bands and metal lines in the $K$
band has been suggested by LW2000: they noted that among empirical spectra with
similarly warm energy distributions those with the strongest H$_2$O
bands tend to have the weakest absorption features of CO and metals
(Na, Ca). All the spectra with an H$_2$O index above 0.6 and
$R-I<2.5$ belong to stars labelled as potentially metal poor by LW2000. 
The construction of pulsating models at subsolar
metallicity and its effect on spectral features (Lan\c{c}on, Scholz \& Wood, 
in preparation)
would help resolve this issue,
and we may expect that intrinsically warmer
models with large pulsation amplitudes will explain some of the
relatively blue spectra with deep H$_2$O absorption.

But metallicity alone cannot account for the discrepancies
in Fig.\,\ref{h2o_vo.fig}. 
This is made clear by the locus of the two individual Miras R Cha
and RS Hya. For these two stars the loci of 
the observed points follow a line that crosses the
locus of the solar metallicity dynamical models.
In the H$_2$O {\em vs.} $R-I$ plot, the effect of metallicity is not that
pronounced as compared to the $V-K$ colour.
Where else could the origin of the discrepancy lie? The water bands
depend on the density and temperature stratifications, but probably
also on the treatment of shocks and on molecular relaxation times,
etc. Apart from metallicity, mild increases of C/O ratios within the
tolerance of M type classification may play some role. 
Potential answers are as numerous as the model ingredients and
only speculations can be made at this stage. Future modeling will
show whether some physical assumption must be modified or whether a
more complete exploration of the model parameters is sufficient.

%--------------------------------------------------------------------
\section{Discussion and Conclusion}
\label{disc_con.sec}
In this paper we have compared the model predicted spectra of 
BSW96 and HSW98 with observations of Miras covering the wavelengths from
0.5 to 2.5\,$\mu$m. For the three Miras with periods similar to those
of the models, namely RS Hya, R Cha and CM Car, the energy distributions 
over the entire spectral range are reproduced reasonably
well, considering
(i) the observational uncertainties and (ii) the large number of relevant
model parameters and the limited number of models actually available.
In many cases, the overall shape is modeled similarly well
for stars with different periods. General agreement is also found between the 
loci of the empirical and synthetic data points in plots 
combining two standard broad band colours. This may indicate that
period is not the most important parameter in determining the range of 
resulting spectrophotometric properties.

When investigating specific spectral features, it is not possible
to fully assess which discrepancies between models and observations
are simply the consequence of the limited sample of available synthetic 
spectra. Discrepancies are seen in the slope of the
spectra between 1 and 1.3\,$\mu$m. Most models display a flat
continuum whereas the observed spectra show both blue and red
slopes in this wavelength region. Models close to maximum (phase 0.8)
do account for the blue slopes. New 0.8 phase models of
the P series were essential to fit to a number of the observed
spectra. In a few cases, models with dust produce red slopes in
this 1\,$\mu$m region and fit better to the observed spectra.
However, not all the spectra having red slopes are well
represented by the limited available dusty models. It is also worth
noting that, when dust helps achieving good agreement for
one spectrum of a star, dust-free models provide satisfactory
fits at other phases of the same star. This may indicate a phase
dependence of dust formation. But a detailed phase-correlated
study, involving exhaustive dust modeling, needs to be done 
before arriving at any definitive conclusion. Missing molecular opacities
in the 1\,$\mu$m region might also be a cause for disparities
at those wavelengths.

The agreement between observed and model-predicted molecular
bands is far from perfect. For VO (1.05\,$\mu$m), the range of 
predicted index values is consistent with observations,
but it is seen that for red $V-K$ or $R-I$ colours
the models produce weaker VO indices than are observed. 
For H$_2$O, the models predict very strong
absorption bands for the reddest spectra. None of the observations 
of LW2000 reach these extreme values of the H$_2$O absorption index.
At the blue end of the colour distribution, on the contrary,
the observed H$_{2}$O bands are stronger than the model-predicted ones. 
On one hand, problems of completeness and of approximations of
opacities used in the models (BSW96, HSW98 and Brett 1990) may be
significant. These are essentially mean opacities rather than
extinction coefficients based on an opacity distribution function or
an opacity sampling technique, neither of which can be applied
directly to a dynamic atmosphere with pronounced velocity
stratification (cf. Scholz 2003; for recent approaches to opacity
distribution functions see Baschek et al. 2001, Wehrse 2002). For
H$_{2}$O absorption, the empirical data of Ludwig (1971) were
adopted. On the other hand, features such
as the H$_2$O band depths and shapes are sensitive to the
precise structure of the outer layers of the atmospheres, which 
might have to be further improved.
In some of the studied models (in particular P18, P38, see the Appendix), the  
shape of the bands clearly shows the combination of emission and absorption
components. Such a phenomenon was noted in observations
at longer wavelengths (Yamamura et al. 1999), but is not obvious in the 
near-IR spectra of LW2000. Because the H$_2$O bands carry the 
potential of revealing the extended atmospheric structure and, thus,
improving the interpretation of stellar diameter measurements
(TLS03), it will be important to continue to invest
effort into their precise modeling.

Finally, our study underlines the importance of using data with 
a broad wavelength coverage when testing the validity of the
models. Any restriction of the observed range makes fits 
easier to achieve and thus less constraining.
Currently, the determination of fundamental stellar parameters 
from spectral fits remains difficult even when optical and
near-IR spectra are available simultaneously. Fits result
in a rather wide collection of equally
acceptable sets of fundamental stellar parameters.
One reason for
this is that the most common parameters, effective temperature,
mass, luminosity, metallicity are not sufficient to describe the
structure of a Mira atmosphere\,; hidden parameters that provide
a more precise description may have effects that compensate 
each other partly, leading to degeneracies. Other reasons
are to be found in the `primitive' nature of the models 
and the restricted phase-cycle-parameter coverage. 
Indeed, it is the quality of the `best fits' between the available models
and observations that in the end also defines how tightly 
the model rejection criteria can be set. Improved 
parameter coverage and model input physics should lead to better fits
to the spectral data, and thus enable us to discriminate between model
parameters more radically. Along with this, increased wavelength
coverage in the data and enhanced spectral resolution of
certain features would help in constraining the parameter
estimates. As an example, one might investigate the relative
strengths of OH and CO lines in the 1.5 -- 1.7$\,\mu$m window,
which differ significantly from one Mira to the other (LW2000).
High-accuracy, multibaseline, near-continuum
interferometric observations covering a large range of phase-cycle
will help in retrieving the monochromatic disk brightness
distribution which gives the only direct information regarding
the geometry and physics of a Mira and constrain the stellar
parameters by defining the proper atmospheric stratification.
Achieving this and improving the situation will also be a task 
of future observational and theoretical work.

%--------------------------------------------------------------------
\begin{acknowledgements}
This research was supported in part by the Australian Research Council and
the Deutsche Forschungsgemeinschaft within the linkage project `Red Giants',
and by a post-doctoral fellowship of the french Minist\`ere de la Recherche.
\end{acknowledgements}
%--------------------------------------------------------------------

%------------------------------------------------------------------------------
\appendix

\section{Parameters for the time series of the models.}

In Table\,\ref{param.tab}, we list the parameters of the time
series of the Mira models. This is an extract of Table 3 of BSW96 (Z,
D, E series) and Table 2 of HSW98 (P, M, O series). 
Newly included are the phase 0.2 and 0.8 models of the P and the M series. 

\begin{table}
\caption{Parameters of the time series of the new Mira models.
The columns: visual phase $\phi_{vis}$; luminosity $L$;
Rosseland radius $R$; 1.04 near-continuum radius $R_{1.04}$;
effective temperatures $T_{\rm eff}$ and $T_{1.04}$).
The model names consist of a
letter defining the series followed by the pulsation phase $\times$ 10.}
\label{param.tab}
\begin{tabular}{lllllll}
\hline
Mod. & $\phi_{vis}$ & $L$ & $R$  & $R_{1.04}$ &
$T_{\rm eff}$ & $T_{1.04}$ \\
 & & ($L_{\odot}$) & ($R_{\rm p}$) & ($R_{\rm p}$) &(K) &(K) \\
\hline
Z08 & 0+0.8 & 8140 & 0.90 & 0.90 & 3770 & 3760\\
Z10 & 1+0.0 & 7650 & 1.10 & 1.11 & 3350 & 3340\\
Z12 & 1+0.24& 6860 & 1.13 & 1.13 & 3220 & 3220\\
Z15 & 1+0.5 & 3860 & 0.89 & 0.89 & 3150 & 3140\\
Z18 & 1+0.8 & 8230 & 0.90 & 0.90 & 3770 & 3770\\
Z20 & 2+0.0 & 7750 & 1.11 & 1.12 & 3350 & 3340\\
Z22 & 2+0.24& 6840 & 1.13 & 1.13 & 3220 & 3210\\
Z25 & 2+0.5 & 3830 & 0.89 & 0.89 & 3140 & 3140\\[2mm]

D08 & 0+0.8 & 3500 & 0.90 & 0.90 & 3050 & 3050\\
D10 & 1+0.0 & 4490 & 1.04 & 1.04 & 3020 & 3020\\
D12 & 1+0.2 & 4920 & 1.09 & 1.10 & 3010 & 3010\\
D15 & 1+0.5 & 2210 & 0.91 & 0.90 & 2710 & 2720\\
D18 & 1+0.8 & 3510 & 0.90 & 0.90 & 3050 & 3050\\
D20 & 2+0.0 & 4560 & 1.04 & 1.05 & 3030 & 3020\\
D22 & 2+0.2 & 4760 & 1.09 & 1.09 & 3000 & 2990\\
D25 & 2+0.5 & 2170 & 0.91 & 0.90 & 2690 & 2700\\[2mm]

E08 & 0+0.83 & 4790 & 1.16 & 1.07 & 2330 & 2440\\
E10 & 1+0.0  & 6750 & 1.09 & 1.09 & 2620 & 2630\\
E11 & 1+0.1  & 8780 & 1.12 & 1.11 & 2760 & 2770\\
E12 & 1+0.21 & 7650 & 1.17 & 1.15 & 2610 & 2640\\[2mm]
%\hline
%\end{tabular}
%\end{table}
%
%\begin{table}
%\caption{(continued; will be one table in final
%print)}
%\begin{tabular}{lllllll}
%\hline
%Mod. & $\phi_{vis}$ & $L$ & $R$  & $R_{1.04}$ &
%$T_{\rm eff}$ & $T_{1.04}$ \\
% & & ($L_{\odot}$) & ($R_{\rm p}$) & ($R_{\rm p}$) &(K) &(K) \\
%\hline
P05 & 0+0.5 & 1650 & 1.20 & 0.90 & 2160 & 2500\\
P08 & 0+0.8 & 4260 & 0.74 & 0.74 & 3500 & 3500 \\
P10 & 1+0.0 & 5300 & 1.03 & 1.04 & 3130 & 3120\\
P12 & 1+0.23& 4540 & 1.38 & 1.30 & 2610 & 2680\\
P15 & 1+0.5 & 1600 & 1.49 & 0.85 & 1930 & 2560\\
P18 & 1+0.8 & 4770 & 0.77 & 0.77 & 3520 & 3520 \\
P20 & 2+0.0 & 4960 & 1.04 & 1.04 & 3060 & 3060\\
P22 & 2+0.18& 4400 & 1.32 & 1.26 & 2640 & 2700\\
P25 & 2+0.5 & 1680 & 1.17 & 0.91 & 2200 & 2500\\
P28 & 2+0.8 & 5200 & 0.79 & 0.79 & 3550 & 3550 \\
P30 & 3+0.0 & 5840 & 1.13 & 1.14 & 3060 & 3050\\
P35 & 3+0.5 & 1760 & 1.13 & 0.81 & 2270 & 2680\\
P38 & 3+0.8 & 5100 & 0.78 & 0.78 & 3570 & 3570 \\
P40 & 4+0.0 & 4820 & 1.17 & 1.16 & 2870 & 2880 \\[2mm]

M05 & 0+0.5 & 1470 & 0.93 & 0.84 & 2310 & 2420\\
M08 & 0+0.75 & 4780 & 0.81 & 0.81 & 3320 & 3320 \\
M10 & 1+0.0 & 4910 & 1.19 & 1.18 & 2750 & 2760\\
M12 & 1+0.26& 2990 & 1.33 & 1.12 & 2300 & 2500\\
M15 & 1+0.5 & 1720 & 0.88 & 0.83 & 2460 & 2530\\
M18 & 1+0.75 & 4840 & 0.81 & 0.81 & 3310 & 3310 \\
M20 & 2+0.0 & 4550 & 1.23 & 1.20 & 2650 & 2680\\
M22 & 2+0.27& 2850 & 1.27 & 1.10 & 2330 & 2490\\[2mm]

O05 & 0+0.5 & 5020 & 1.12 & 1.00 & 2050 & 2130\\
O08 & 0+0.8 & 4180 & 0.93 & 0.91 & 2150 & 2170\\
O10 & 1+0.0 & 7070 & 1.05 & 1.01 & 2310 & 2360\\
\hline
\end{tabular}
\end{table}

The P and M models at phase 0.8 
are particularly warm and compact, a phenomenon already found in the
Z and D fundamental-mode pulsation series. 
As the new shock front emerges from deep layers, densities
are fairly low in the outer and middle parts of the atmosphere and
unity continuous optical depth is reached at small distances from the
star's centre.  The star appears small in continuum bandpasses and effective
temperatures are high. The effect is particularly impressive for the self-excited
HSW98 model series for which differences between pre-maximum and maximum
continuum radii are of the order of 0.3 parent star radii, 
resulting in dramatically higher $T_{\rm 1.04}$ effective temperatures 
around phase 0.8 than at maximum despite similar luminosities.
The spectra have relatively
blue colours (bluer than at phase 0.0 or 0.5), but significant
molecular features nevertheless. The blue slope between
1 and 1.3\,$\mu$m is particularly impressive for models
P18 and P38. The remarkably irregular light curve of the P model
series (see HSW98) shows that in these two cycles maximum light
was actually reached before the phase labelled 0 (see
Sect.\,\ref{compare.sec} for uncertainties of phase adjustments). 

\begin{figure}
\includegraphics[clip=,width=0.5\textwidth]{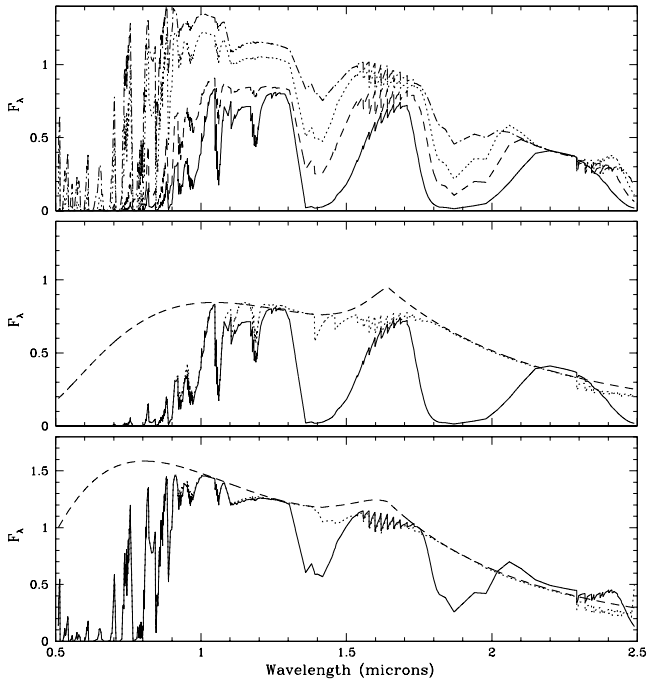}
\caption{Molecular emission and absorption in models from the P series.
The top panel shows 4 models, normalized at 2.2\,$\mu$m to
highlight differing shapes in that region.
P15 - solid; P18 - dotted; P20 - short dash; P38 -
dot-short dash. The two lower panels compare the
model spectra of P15 (middle) and P18 (bottom) with the
pure continuum spectrum (short dash) and spectrum without H$_{2}$O
absorption (dotted) without any normalization. The
continuum spectrum is obtained by setting all molecular opacities
to zero after the model atmosphere has been constructed with the
complete opacity list, and for the spectrum without H$_{2}$O only the
opacities due to H$_{2}$O are switched off.}
\label{apend.fig}
\end{figure}

Models P18 and P38 also stand out because of the ``peculiar" shape of their
water bands. In the wings of what usually appears as a broad absorption
band, i.e. around 2.05 and 2.4\,$\mu$m, bumps that could be described
as ``emission shoulders" are present (top of Fig.\,\ref{apend.fig}).  
In the H window, these ``shoulders" are not as obvious but manifest only 
as a modification of the curvature of the resulting energy distribution.
This phenomenon has never been mentioned in work based on spectroscopy
below 2.5\,$\mu$m but has been suggested to occur at longer wavelengths
(Yamamura et al. 1999, Matsuura et al. 2002). The middle panel of 
Fig.\,\ref{apend.fig} shows that excess emission above the
pure continuum level can sometimes occur in the K band in a way that would
be extremely difficult to identify in low resolution spectra.
  
\end{document}